\documentclass[12pt]{article}\usepackage[hyperfootnotes=false]{hyperref}
\usepackage{epsfig}
\usepackage{float}
\usepackage{dsfont}

\usepackage{bbold}

\usepackage{caption}
\usepackage{subcaption}

\usepackage{amsmath}
\usepackage{amssymb}
\usepackage{graphicx}
\newlength{\abstractwidth}
\renewcommand{\thefootnote}{\fnsymbol{footnote}}
\renewcommand{\thanks}[1]{\footnote{#1}} % Use this for footnotes
\newcommand{\starttext}{
\setcounter{footnote}{0}
\renewcommand{\thefootnote}{\arabic{footnote}}}

\newcommand{\be}{\begin{equation}}
\newcommand{\bea}{\begin{eqnarray}}
\newcommand{\eea}{\end{eqnarray}}
\newcommand{\beq}{\begin{equation}}
\newcommand{\ee}{\end{equation}}

\def\simleq{\; \raise0.3ex\hbox{$<$\kern-0.75em
\raise-1.1ex\hbox{$\sim$}}\; }
\def\simgeq{\; \raise0.3ex\hbox{$>$\kern-0.75em
\raise-1.1ex\hbox{$\sim$}}\; }

\def\bi{\begin{itemize}}
\def\ei{\end{itemize}}

\def\t{\tau}

\def\bn{\bigskip \noindent}

\makeatletter
\g@addto@macro\normalsize{%
  \setlength\abovedisplayskip{10pt}
  \setlength\belowdisplayskip{20pt}
  \setlength\abovedisplayshortskip{10pt}
  \setlength\belowdisplayshortskip{20pt}
}
\makeatother

\usepackage{color}

\renewcommand{\title}[1]{\vbox{\center\LARGE{#1}}\vspace{5mm}}
\renewcommand{\author}[1]{\vbox{\center#1}\vspace{5mm}}
\newcommand{\address}[1]{\vbox{\center\em#1}}

\begin{document}
  
\begin{titlepage}

\rightline{}
\bigskip
\bigskip\bigskip\bigskip\bigskip
\bigskip

\centerline{\Large \bf { Why do Things Fall?}}

\bn
\bn

\bn

\bigskip

\bigskip
\begin{center}

\author{Leonard Susskind}

\address{Stanford Institute for Theoretical Physics and Department of Physics, \\
Stanford University, Stanford, CA 94305-4060, USA}

\end{center}

\begin{center}
\bf     \rm

\bigskip

%\vspace{1cm}
\end{center}

\begin{abstract}

This is the first of several short notes in which I will describe phenomena that
illustrate GR=QM. In it I explain that the gravitational attraction that a black
hole exerts on a nearby test object is a consequence of a fundamental law of quantum
mechanics---the tendency for complexity to grow. 
\medskip
\noindent
\end{abstract}

\end{titlepage}

\starttext \baselineskip=17.63pt \setcounter{footnote}{0}

\vfill\eject

\vfill\eject

\section{Introduction}
There are a number of  correspondences between quantum mechanics and gravity that are suggestive of a much deeper connection than we might have imagined several years ago \cite{Susskind:2017ney}. ER=EPR  is one \cite{Maldacena:2013xja}, and the relation between the generic growth of quantum complexity and the expansion of space behind the horizons of black holes is another \cite{Brown:2017jil}. There are more that I will publish in a series of short notes. 
In this first note I will point out  a relation between ordinary gravitational attraction and the general properties of quantum chaos\footnote{An earlier version of this paper claimed that the chaos bound of Maldacena, Shenker, and Stanford could be related to the Einstein causality bound. The argument was wrong and has been removed from this version.}. 

\section{Momentum and Operator Size}

Why do things fall toward the horizon of a black hole? The usual answer is: Gravity makes them do it. But if, as I have suggested, GR=QM \cite{Susskind:2017ney} then there should be another explanation of purely quantum origin. So what I propose is that there is a quantum-information meaning to gravitational attraction.
In this note I will argue that the gravitational attraction is nothing but the statistical tendency for complexity (manifested by \it operator size\rm) to grow in chaotic quantum systems\footnote{There is  a similarity with E. Verlinde's entropic theory of gravity \cite{Verlinde:2010hp}, but in  this paper the growth of complexity and  growth of operator size play the leading role.}. Gravitational attraction is a manifestation of the tendency for complexity to increase: the second law of complexity \cite{Brown:2017jil}.

\subsection{Note About Time Coordinates}
In what follows I will sometimes use the Schwarzschild time coordinate in which the metric has the usual form,
$$
-f(r) dt^2 + f(r)^{-1}dr^2 +r^2 d\Omega^2
$$
and also the Rindler time in which the near horizon metric has the form
$$
-\rho^2 d\tau^2 + d\rho^2 + R_s^2 d\Omega^2.
$$
Here $\rho$ is proper distance from the horizon, $\tau$ is Rindler time, and $R_s$ is Schwarzschild radius.
The Rindler time $\tau$ is a dimensionless hyperbolic boost angle. 

Schwarzschild  and Rindler times are proportional to each other with a proportionality factor that depends on the temperature of the black hole.

\subsection{Size}

The idea of a  momentum-dependent size is not  a new idea; it dates back to the late 1960's, appearing in both Feynman's parton theory and the earliest work on string theory. According to Feynman the number of partons in a hadron depends on its momentum with more and more ``wee partons" appearing as the particle's momentum increases. 
 Similarly, the size of a relativistic string grows with momentum. In \cite{Karliner:1988hd}\cite{Susskind:1993aa} it was observed that
 the length of a string (measured along the string itself) grows proportional to its momentum. The growth is due the time dilation as the velocity of the center of mass of the string increases. More and more oscillating modes of the string become visible as the velocity becomes relativistic and that in turn causes the length of string to grow \cite{Mezhlumian:1994pe}.
If the string is made of partons (or string-bits), as long advocated by Thorn \cite{Thorn:1996fa}, the number of bits will grow with momentum. If the string is a gauge-theory string the number of gauge quanta grows. 

BFSS Matrix theory is a precise version of the size-momentum relation in which particles are composites of D0-branes with the size  (number of D0-branes), exactly equaling the linear momentum in units of a light-like compactification radius.

In the modern theory of gauge-gravity duality the size-momentum connection reappears in the theory of precursor operators. In the presence of a black hole a particle may be injected into bulk space by applying a simple operator $W(0)$ such as a single trace operator smeared so that it carries a thermal momentum. That particle will accelerate  inward toward the horizon. At the same time the operator $W$ will evolve to $W(t) = e^{-iHt}W(0)e^{iHt}$. Expanding $W(t)$ in the fundamental fields of the theory one finds that the average number of gauge quanta in $W(t)$ grows exponentially with time, thus defining a time-dependent size. The hypothesis of this paper is that  size is dual to  radial momentum, and that the operator growth is dual to the acceleration due to gravitational attraction toward the black hole. 

Let us consider a test particle near an AdS black hole. The angular position of the particle  can be described in terms of  angular momentum,  an exactly conserved quantity in the gauge theory; wave packets of well defined angular position are superpositions of angular momentum states. By contrast the radial position and momentum are encoded in a much less obvious way. There is no symmetry of the CFT corresponding to radial momentum and it is not obvious that the gauge/gravity dictionary contains a precisely defined radial momentum. Nevertheless, in some approximation, at least when semiclassical Einstein gravity is a good approximation, some notion of the radial momentum of a particle should exist. More precisely the average momentum of an infalling wave packet should exist.

Let us suppose that the particle is created by applying a smeared time-reversal-invariant CFT field operator, $W$, at Rindler time $\t=0.$ The time-reversal invariance means the particle would be at rest at $\t=0.$ Now let us go to a later time $\t.$ From the standpoint of an observer at that later time the particle was created by a ``precursor" operator,
\be 
W(-\t) = e^{iH\t}We^{-iH\t}.
\label{W(t)}
\ee

It is expected that for any system that behaves like a large black hole the size grows like \cite{Kitaev}\cite{Size}\cite{Roberts:2018mnp}
$$ 
size \sim e^{\lambda t}
$$
with $\lambda$ being the limiting Lyapunov exponent 
\be 
\lambda = \frac{\beta}{2\pi},
\ee
Thus for all large black holes,
\be 
size \sim e^{\frac{2\pi}{\beta} t}
\ee

\subsection{Momentum}
 To see how the momentum grows we will consider the motion of a classical particle falling toward the horizon.
The near-horizon Rindler-like geometry of non-extremal black holes has  (time-like) metric,
\be 
d\sigma^2 = \rho^2 d\tau^2 - d\rho^2 + R_s^2 d\Omega^2.
\ee
Let us consider a massive particle moving in the radial direction along the trajectory $\rho(\tau)$. The Lagrangian for the particle is 
\be 
L= -m \sqrt{\rho^2 -\dot{\rho}^2}
\ee
where dot means derivative with respect to $\tau.$
The momentum conjugate to proper distance  $\rho$ is,
\be  
P = \frac{m\dot{\rho}}{ \sqrt{\rho^2 -\dot{\rho}^2} }.
\label{P}
\ee

\begin{figure}[H]
\begin{center}
\includegraphics[scale=.25]{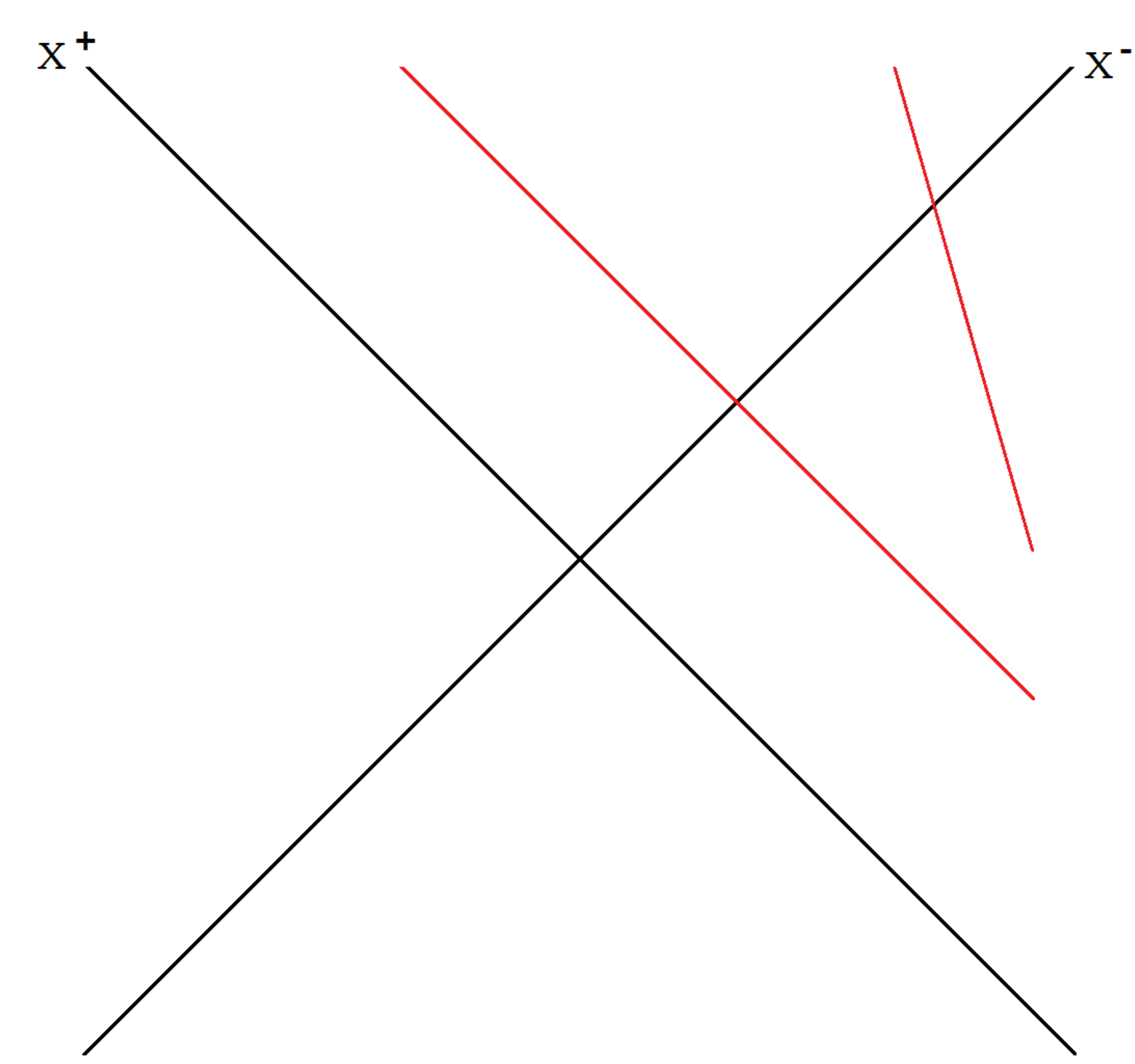}
\caption{}
\label{rindler-fall}
\end{center}
\end{figure}

Figure \ref{rindler-fall} shows two trajectories that fall through the horizon, one light-like and one time-like. 
\bn
The time like trajectory has the form,
\be 
X^+ = a X^- +b 
\ee
with $a>0.$ Using \ref{P} the magnitude of the radial momentum is,
\be 
P = m\frac{e^{\t} - a e^{-\t}}{\sqrt{2a}},
\ee
which asymptotically grows like,
\be
|P| = P_0 e^{\tau}
\label{P(t)}
\ee

For the light-like trajectory we have to do the usual limit in which $m\to 0$ as $a\to 0,$ keeping the momentum finite. In that case $P$  grows precisely like $e^{\t}$
This property \ref{P(t)} of infalling point particles is universal and holds for all black holes in all dimensions, as long as they are not extremal.
The question is what quantity in the QFT description represents the radial momentum?

I'll now propose the following correspondence: The magnitude of the average infalling radial momentum of an object is the size of the precursor $W(t).$  From this correspondence and \ref{P(t)} we conclude that the size grows like $e^{\tau}$.
It is therefore very interesting that the size of operators in chaotic theories---SYK being an example---grows exponentially in a characteristic way \cite{Size}. The growth is  described by a Lyapunov exponent  \cite{Kitaev}  called $\lambda,$
\be 
\rm size\it \sim e^{\lambda t}.
\label{Lyap}
\ee

Now the punch line: The time in \ref{Lyap} is the boundary gauge theory time which is universally related to the Rindler time by,
\be 
t =  \frac{\beta}{2\pi} \tau
\label{beta over 2 pi}
\ee
where $\beta$ is the inverse temperature of the black hole.
Thus the exponential increase of momentum of the falling particle, $P \sim e^{\t}$, translates to a growth of size,
\be 
\rm size\it \sim e^{(2\pi/\beta) t}
\label{size2pioverbeta}
\ee
or in terms of the Lyapunov exponent,
\be 
\lambda = \frac{2\pi}{\beta}
\label{L2pioverbeta}
\ee

This exactly saturates the chaos bound of Maldacena, Shenker, and Stanford, and moreover it is the expected behavior of precursor-size in the environment of a black hole. Thus we see a precise parallel between  operator growth for  strongly coupled chaotic quantum systems and the phenomena of acceleration in a gravitational field.

\section{Corrections to the Lyapunov Exponent }

Equations \ref{size2pioverbeta} and \ref{L2pioverbeta} indicate that black holes saturate the fast-scrambling bound of \cite{Maldacena:2015waa} but seem to preclude the possibility that some black holes have Lyapunov exponents less than $2\pi/\beta.$ However, Shenker and Stanford  \cite{Shenker:2014cwa} have shown that string-theoretic corrections to $\lambda$ may decrease its value. The resolution of this tension can be seen from the form of the corrections. The change of the Lyapunov exponent found by Shenker and Stanford is,
\be 
\Delta \lambda \sim - \frac{2\pi}{\beta}%\left(
\frac{l_s^2}{l_{AdS}^2}
%\right)
\ee
where $\l_s$ is the string length scale.

For black holes in flat space this would be replaced by
\be 
\Delta \lambda \sim - \frac{2\pi}{\beta}%\left(
\frac{l_s^2}{R_s^2}
%\right)
\ee

 This means that the correction is non-negligible \it only \rm if 
the string length scale is comparable to the black hole Schwarzschild radius. In other words the Lyapunov exponent is corrected only to the extent that the geometric description of the black hole horizon breaks down. In the terms of 
\cite{Susskind:1993ws} \cite{Horowitz:1996nw} gravity is not strong enough to completely make the transition from string-like behavior to black hole behavior. In this respect it would be interesting to study the chaotic and scrambling properties of weakly coupled thermally excited string states.

\bn

\bn

\end{document}